\documentclass[11pt]{article}
\usepackage{moriond,epsfig}

\def\be{\begin{equation}}
\def\ee{\end{equation}}
\def\bea{\begin{eqnarray}}
\def\eea{\end{eqnarray}}

\begin{document}
\vspace*{4cm}
\title{COSMOLOGICAL OBSERVATIONS IN THE RADIO DOMAIN: THE CONTRIBUTION OF EXTRAGALACTIC SOURCES}

\author{A.TARTARI, M.ZANNONI, M.GERVASI, G.SIRONI $\&$ S.SPINELLI}

\address{Dip. di Fisica "G.Occhialini" - Universit\`{a} degli Studi di Milano-Bicocca \\
Piazza della Scienza, 3 - 20126 Milano, Italy}

\maketitle\abstracts{ The low frequency tail of the CMB spectrum,
down along the radio range ($\sim$1 GHz), may carry weak spectral
distortions which are fingerprints of processes occurred during
different epochs of the thermal history of the Universe, from $z\sim
3\times 10^6$ to reionization. TRIS and ARCADE2 are the most recent
experiments dedicated to the exploration of this chapter of CMB
cosmology. The level of instrumental accuracy they reached in the
determination of the absolute sky temperature is such that the
removal of galactic and extra-galactic contamination is the true
bottleneck towards the recovery of the cosmological signal. This
will be certainly the case also for future experiments in the radio
domain. Here we present an update of a study originally done to
recognize the contribution of unresolved extra-galactic radio
sources to the sky brightness measured by TRIS. Despite the specific
context which originated our analysis, this is a study of general
interest, improved by the inclusion of all the source counts
available up-to-date from 150 MHz to 8.4 GHz.}

\section{Scientific case}

The frequency spectrum of the Cosmic Microwave Background (CMB)
recovered by the FIRAS istrument~\cite{fixsen1} on board the COBE
satellite  is an almost perfect planckian spectrum at the
thermodynamic temperature of $2.728\pm 0.004 $ K, even if a more
recent analysis and an update of this result can be found in a
recent work by Fixsen~\cite{fixsen2}. No signatures of spectral
distortions have been found on the monopole scale. Nevertheless,
there are important arguments~\cite{ZS} suggesting that deviations
from a blackbody curve originated in the pre-recombination Universe,
may be present in the CMB spectrum. In particular, we know that if
an energy injection in the photon-baryon fluid occurred in the
redshift range $\sim 10^5 < z < \sim 3\times 10^6$, corresponding to
a temperature of the photon-baryon fluid $\sim 0.1$keV $< T < \sim
1$keV, than the system relaxes towards a kinetic equilibrium, the
photons having Bose-Einstein spectrum with a chemical potential $\mu
\ne 0$. Several mechanisms have been proposed as possible sources of
perturbation, among them decay of relic massive
particles~\cite{HuSilk}, dissipation of primordial magnetic
fields~\cite{jedamzik}, annihilation of relic
particles~\cite{McDonald} and dissipation of acoustic oscillations
in the fluid~\cite{barrow}. Was the spectrum a Bose-Einstein one, we
would observe a dip in the plot Temperature \emph{vs} Frequency of
the CMB at frequencies $<\sim$1 GHz, given the present estimate of
$H_0$ (Hubble parameter) and $\Omega _b$ (baryon density). The
amplitude of this distortion, $\Delta T/T\simeq
\mu(\Omega_bh^2)^{-2/3}$ ($h$ being $H_0$ in units of 100
(km/s)/Mpc), is strongly model dependent. Extrapolating to lower
frequencies the FIRAS upper limit on $\mu$, we expect a temperature
dip smaller than few tens of mK.

Another effect able to produce a monopole scale spectral distortion
is free-free emission during reionization. This effect produces a
signal $\Delta T/T\propto Y_{ff}\lambda^2$, which is essentially the
optical depth to bremsstrahlung, $\lambda$ being the wavelength and
$Y_{ff}$ a distortion parameter as defined by Bartlett and
Stebbins~\cite{bartlett}. Despite new tighter limits on $Y_{ff}$
obtained by Gervasi et al.~\cite{gervasi1}, we are still far from
the possibility of testing reionization scenarios like those
investigated by Weller et al.~\cite{weller}.

Both effects are more relevant at decimetric wavelengths, where (1)
calibrations are more difficult due to the size of antennas, (2)the
brightness of the Galaxy overcomes the CMB and (3) the signal of
Unresolved Extragalactic Radio Sources (UERS), the subject of this
study, may introduce a temperature offset if not properly evaluated
and subtracted.

Two recent experiments searched for low frequency CMB spectral
distortions: TRIS~\cite{zannoni} and ARCADE2~\cite{arcade}. In the
framework of TRIS, a new estimate of the brightness temperature of
UERS has been proposed by Gervasi et al.~\cite{gervasi2}. Here we
present an update of those results, exploiting all the most recent
radio source counts from 150 MHz to 8.4 GHz and adding a new
frequency to those used in the previous study.

\section{Unresolved Extragalactic Radio Sources contamination}

\subsection{Aims and Methods of this study}

As stated in the previous section, the purpose of this work is to
develop the earlier work by Gervasi et al.~\cite{gervasi2} enlarging
the data-set described in table 1 of this reference. Here we
included all the deepest radio source counts done in the last few
years, especially those obtained at VLA and GMRT. In particular, we
enriched our analysis taking into account new results at 150
MHz~\cite{ishwara}, 610 MHz~\cite{garn} and 1400 MHz~\cite{ibar}. We
completed the 8400 MHz counts by adding the results by Henkel and
Partridge~\cite{henkel}.

\noindent We added also a new frequency in our analysis, namely the
325 MHz channel (Oort et al.~\cite{oort}, Owen et al.~\cite{owen}).

\noindent Typical experiments looking for spectral distortions in
the CMB spectrum have beams $>7^\circ$ (e.g. FIRAS), so that, on
average, pointing in different directions, the radiometer will
detect the same blend of AGNs, quasars and normal galaxies, with
only a poissonian fluctuation in their number. Therefore, UERS are
seen as an isotropic diffuse radiation. We calculate its temperature
in two steps. First, we fit the differential source counts
normalized to the euclidean counts, that is $Q(S)=S^{2.5}dN/dS$, $S$
being the flux. Then we use the definition of brightness
temperature,

\begin{equation}
T_{b,UERS}=\frac{\lambda^2}{2k_B}\int_{S_{min}}^{S_{max}}Q(S)S^{-3/2}dS
\end{equation}

\noindent ($k_B$ Boltzmann constant), to calculate the integrated
contribution of sources between two limiting values of flux. Here a
problem is evident: given the fact that a survey will be complete at
a flux limit $S_{min}$, how can we take into account the
contribution of sources fainter than that? If we stop our
integration within the data range, we get a lower limit of the
temperature of UERS, introducing a bias in our analysis. A solution
is to extrapolate our integral at fainter fluxes. In absence of a
physical cut-off, to circumvent this problem we simply look for a
functional form of Q(S) such that $T_{b,UERS}$ remains finite when
$S_{min}\rightarrow 0$, avoiding a kind of Olbers'paradox.

\subsection{Radio Source counts}

As originally proposed by Gervasi et al.~\cite{gervasi2} we assume

\begin{equation}
Q(S)=Q_1(S)+Q_2(S)=\frac{1}{A_1S^{\varepsilon_1}+B_1S^{\beta_1}}+\frac{1}{A_2S^{\varepsilon_2}+B_2S^{\beta_2}}
\end{equation}\label{lognlogs}

\noindent where $A_i$,$B_i$,$\varepsilon_i$ and $\beta_i$ ($i=1,2$)
are parameters to be fitted. This analytical description, inspired
by source evolutionary models proposed by Danese, Franceschini and
collaborators (see Danese et al.~\cite{danese} and Franceschini et
al.~\cite{franceschini}), has the property of being integrable at
faint fluxes, even if it is not good as log(S) polynomials to
describe features within the experimental range. One of the basic
assumptions underlying our approach is that the four parameters of
$Q(S)$ giving the slopes ($\varepsilon_i$ and $\beta_i$) are
frequency independent. Moreover, following again a suggestion of the
model by Franceschini et al., and analyzing the data at 600, 1400
and 5000 MHz (frequencies where both faint and strong fluxes are
well sampled), we fix $\varepsilon_1=\varepsilon_2$, simplifying
further the description of source counts. Now, the $Q_1$ and $Q_2$
terms, following the notation of Gervasi et al.~\cite{gervasi2}, are
dominant respectively in the strong and faint flux regimes, so that
two ratios can be calculated at the three frequencies with the
widest flux coverage: $r_A=A_2/A_1$ and $r_B=B_2/B_1$, at 600, 1400
and 5000 MHz. These ratios, $r_A$ in particular, can be used to
reconstruct the faint flux tails of counts at those frequencies
where data are not deep enough. Since they show a weak frequency
dependence (see tab.~\ref{tab:rat} for $A_2/A_1$), our choice has
been to use the $r_A$ and $r_B$ calculated at the frequency closest
to the one we want to reconstruct.

\begin{table}[h!]
\caption{The measured $A_2/A_1$ ratios.\label{tab:rat}}
\begin{center}
\begin{tabular}{c|c|c|c}
 & 600 MHz  & 1400 MHz & 5000 MHz  \\
\hline \hline
$A_2/A_1$ &  $0.17\pm 0.02$& $0.23\pm 0.01$ & $0.31\pm 0.03$\\
\end{tabular}
\end{center}
\end{table}

\section{Results}

\begin{center}
\begin{figure}[!h]
\rule{5cm}{0.2mm}\hfill\rule{5cm}{0.2mm}\vskip 0.5cm
\psfig{figure=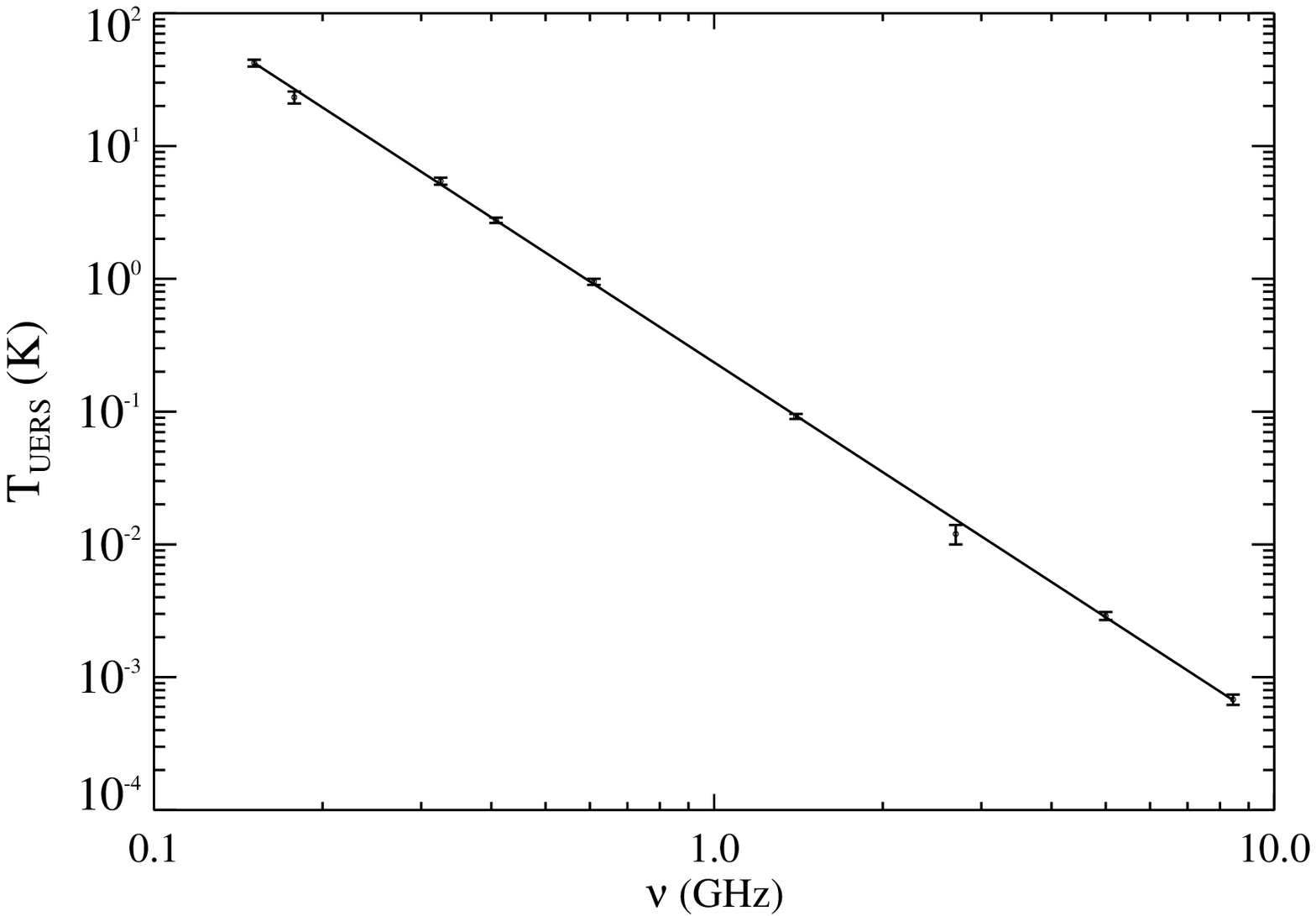,height=55mm}
\psfig{figure=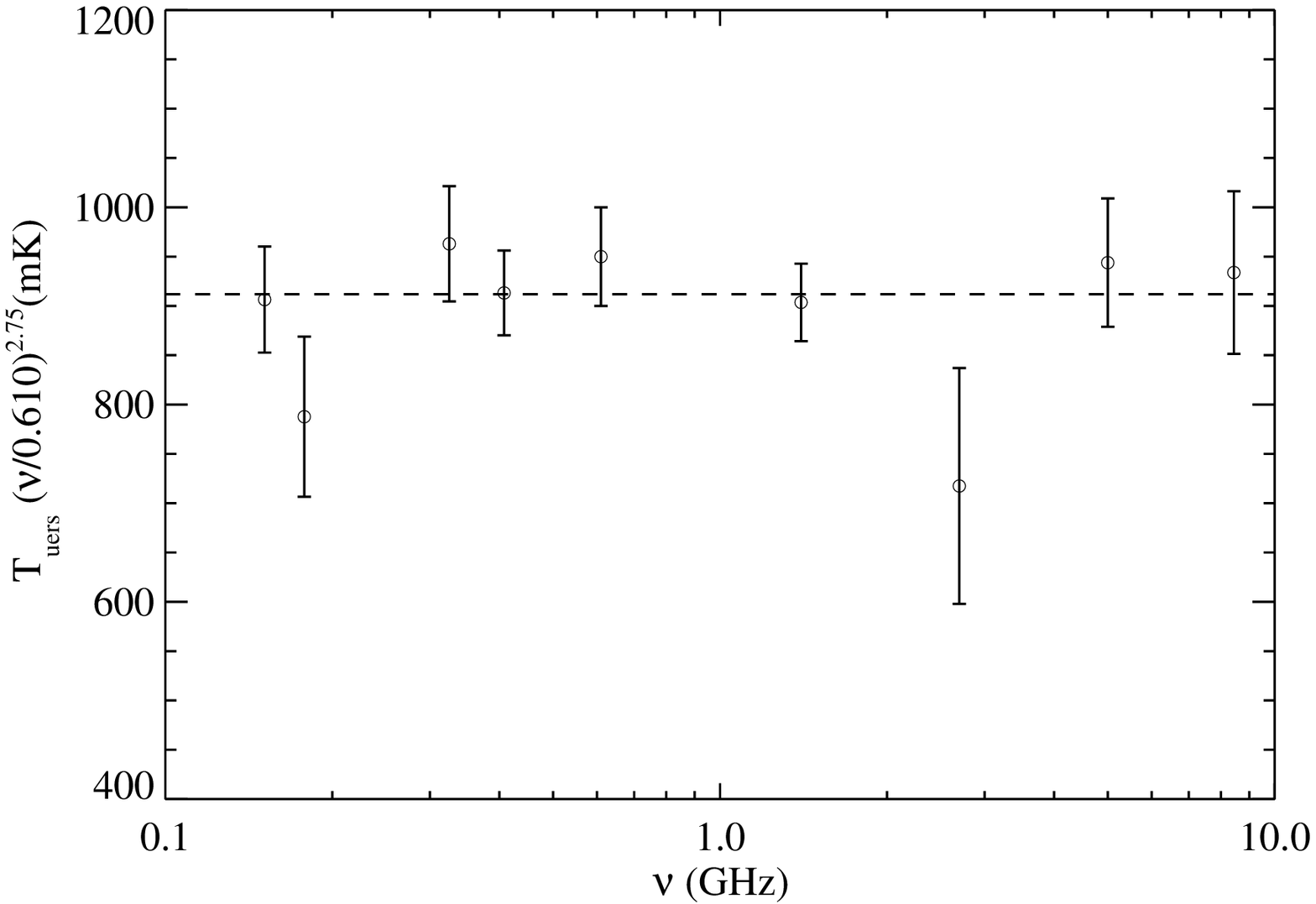,height=55mm}
\caption{Left panel: Temperature vs Frequency plot; calculated
temperatures and power law fit (solid line); 1$\sigma$ error bars.
Right panel: the same as in left panel, but with temperature
rescaled by a factor $\propto \nu^{2.75}$ to appreciate the scatter
of points around the single power law fit.\label{fig:powerlaw}}
\rule{5cm}{0.2mm}\hfill\rule{5cm}{0.2mm}
\end{figure}
\end{center}

The main result we have obtained is a new estimate of the brightness
temperature of the blend of UERS in the range 150 MHz - 8400 MHz.
The calculated temperature values are well described by a single
power law (eq.~\ref{eq:Temp}) with spectral index -2.75 (see
fig.\ref{fig:powerlaw}). Our estimates are well in agreement with
those found by other authors~\cite{dezotti}, and the main result is

\begin{equation}
T_{b,UERS}(\nu)=(0.91\pm 0.02)\Big(\frac{\nu}{610
MHz}\Big)^{-(2.75\pm 0.02)}K
\end{equation}\label{eq:Temp}

\noindent for the contribution of unresolved sources to the overall
temperature of the radio sky.

This formula allows us to calculate the UERS temperature in
correspondence of TRIS frequency channels, namely 600, 820 and 2500
MHz: we find, respectively, $950\pm 20$mK, $408\pm 9$ mK and
$20.6\pm 0.7$ mK.

\section{Conclusions}

This update of our previous work allowed us a better reconstruction
of differential source counts, especially at faint fluxes. In fact,
unlike the first study, we could determine the ratio $A_2/A_1$
finding a hint for weak frequency dependence. Improved source count
fits lead directly to a more accurate estimate of the brightness
temperature: this is very well described by a single power-law
frequency spectrum. Then, using eq.~\ref{eq:Temp}, we could evaluate
the UERS contribution at TRIS frequencies with uncertainties almost
a factor 10 smaller than those commonly assumed before new estimates
by Gervasi et al.~\cite{gervasi2}. By virtue of that, the
subtraction of this foreground is no longer a limiting factor in low
frequency CMB experiments, the overall error budget being now
dominated by the reconstruction of the absolute temperature scale
and by the disentangling of galactic and cosmological signals.
Finally, we conclude that the estimate given by eq.\ref{eq:Temp},
exclusively based on source counts, doesn't agree with the
extragalactic radio excess detected as a diffuse signal by the
ARCADE2 collaboration~\cite{fixsen3}. The discrepancy is around a
factor 5 at 600 MHz, and still the origin of the radio excess
remains mysterious. If its origin is truly extragalactic, there must
be a number of sources able to produce an intense integrated signal,
but individually so weak to escape from detection with radio
interferometers. The case is intriguing and again tells us how non
trivial is searching for CMB spectral distortions, and how
unpredictable are the outcomes.

\end{document}